# Construction Payment Automation Using Blockchain-Enabled Smart Contracts and Reality Capture Technologies

By

Hesam Hamledari & Martin Fischer

# Construction Payment Automation Using Blockchain-Enabled Smart Contracts and Reality Capture Technologies

Hesam Hamledari[1], Martin Fischer[2]

[1] PhD Candidate, Stanford University, Department of Civil & Environmental Engineering, Stanford, CA, United States, hesamh@stanford.edu

[2] Kumagai Professor of Engineering and Professor in Civil & Environmental Engineering, Stanford University, Stanford, CA, United States, fischer@stanford.edu

**ABSTRACT**

This paper presents a smart contract-based solution for autonomous administration of construction progress payments. It bridges the gap between payments (cash flow) and the progress assessments at job sites (product flow). The approach eliminates the reliance on the centralized and heavily intermediated mechanisms of existing payment applications. The construction progress is stored in a distributed manner using content addressable file sharing and in the form of as-built building information models (BIM) and reality capture data; it is broadcasted to a smart contract which automates the on-chain payment settlements and the transfer of lien rights. The method was successfully used for processing payments to 7 subcontractors in two commercial construction projects where progress monitoring was performed using a camera-equipped unmanned aerial vehicle (UAV) and an unmanned ground vehicle (UGV) equipped with a laser scanner. The results show promise for the application of smart contracts and blockchain for automating and decentralizing the transition from job site observations to construction payments.

# 1  INTRODUCTION

Cash flow management is crucial to the financial wellbeing of construction and engineering firms and the on-time and successful delivery of construction and infrastructure projects [1]. Construction progress payments constitute a big portion of cash flow on a project, compensating the general contractor, subcontractors, suppliers, and equipment providers for the work performed during a billing period. The payment practices in the architectural, engineering, and construction (AEC) industry suffer from slow payments [2,3], information asymmetry [4], and low productivity due to ineffective contracting practices [5]. Poor payment practices are a major risk for the industry [6] and create distrust between stakeholders [7].

In the past decade, the AEC industry has experienced an increasing interest in automated progress monitoring at job sites, with research studies focused on reality capture technologies [8]: the use of robotics, artificial intelligence, and building information modeling (BIM) for 1) capturing [9,10], 2) analyzing [11,12], and 3) modeling the job site conditions [13,14]. These efforts result in semantically rich as-built BIMs and progress data that has potential to streamline payments and automate the transition from product flow to cash flow in the construction supply chain.

Despite the availability of digitized progress data, payment automation is far from reality. Projects still rely on traditional payment applications that are time consuming and information intensive [15]. In recent years, digital payment management systems have been developed, commercial solutions that enable project participants to complete the payment application online instead of using paper. This reduces the payment processing effort by 84% [16]. These digital platforms, however, still cannot support payment automation. This is partly due to their inability to use the output of reality capture technologies and partly because they adopt the same underlying manual and intermediated workflows as paper-based applications. As a result, they suffer from centralized control mechanisms and a lack of guaranteed execution [17]. The industry lacks reliable and automated means of translating product flow (job site observations) to cash flow (construction payments); the current workflows are heavily intermediated, manual, and incapable of using reality capture technologies for payment automation.

To address these limitations, this paper introduces a novel smart contract-based solution to automate and decentralize the conditioning of construction payments on the progress assessments, enabled by the on-site deployment of reality capture technologies. It proposes a method for automating the transition from product flow to cash flow, taking advantage of content addressable and distributed data sharing, blockchain's immutability, and the smart contracts' self-enforceability. The method was evaluated in two commercial construction projects where robotic reality capture was used to document the progress at job sites. The proposed solution was successfully utilized to process payments to a series of subcontractors.

## 2 BACKGROUND

There are two common themes to most studies on the applications of smart contract and its underlying technology, blockchain, in AEC industry: 1) increasing transparency via blockchain's traceability feature and 2) creating improved governance models via the use of smart contract's self-executability features.

1. Transparency: the blockchain's traceability feature fosters collaboration among AEC stakeholders [18], and it increases the trust in the data used in construction management applications [19,20]. This increased trust and transparency are argued to be the key features that make the technology appealing to construction practitioners [21-25]. A case study of permissioned and public blockchain finds this value proposition to be particularly of importance in projects where stakeholders are geographically distant, and it is hard to naturally establish trust [26].

One study [27] classified the design events related to BIM-based processes and proposed a blockchain-based framework for keeping an immutable record of these events, enabling project members to review the history of changes to a model. Another work [28] tracked the modifications to BIM by storing the cryptographic hashes of files on the blockchain. A drawback to such approach is the information redundancy. To address this limitation, a semantic differential approach was introduced that allowed for changes to BIM to be recorded instead of entire models [29], reducing the data storage needs.

A framework was proposed to integrate internet of things, smart contracts, and information models to improve the tracking of maintenance and operation records for physical assets [30]. The ability to keep track of changes and the ownership of data was stated to be the major benefit of blockchain for asset information management systems [31]; this motivated a conceptual framework for asset information models based on blockchain and focused on operation stage [32]. These studies are a great step toward creating shared and transparent view of information; however, they need to go beyond the theory and see further validation in real-world projects.

A framework was proposed for tracing the quality information for precast concrete components using blockchain [33]. The passive RFID technology was integrated with blockchain [34] for tracking ready-mix concrete panels from production to on-site delivery; both solutions need to be validated in future. The possibility of creating a single-source of truth was identified as the most important benefit of blockchain in construction supply chain management [34]. A comparison of traditional and blockchain-based solutions reveals that the latter is better suited for tracking the embodied carbon content in the construction supply chain due to 11 features including decentralization, anonymity, and increased security [35].

A case study application of blockchain in water infrastructure projects [36] demonstrated that significant performance improvements can be achieved. The technology was applied on 44 tasks and across 12 real-world wastewater treatment plant projects. The blockchain-based solution was shown to reduce work hours by 49% [36].

2. Governance: new governance models are needed to facilitate blockchain adoption [37], and the lack thereof is the primary barrier to successful use of smart contracts in AEC [38]. Researchers aim to use the smart contract's self-executability to improve the existing governance models or create new ones.

Through a case study-based approach and analyzing the Dubai government's use of smart contracts, a new model was proposed for delivering government services on a blockchain-based ecosystem [39]. A novel method proposed the use of smart contracts for creating a financial system for integrated project delivery (IPD) [40], keeping an immutable record of cost performances against 5D BIM. Non-owner parties interact with smart contract by communicating the performance (retrieved from cost-loaded BIM) and can invoke payments at milestones; the smart contract processes requests based on planned and achieved profits. While the work is yet to be validated, the proof of concept provides potential for creating transparency and eliminating management overhead in IPD.

It was argued [41] that smart contract can automate the collaborations across the construction value chain by reducing the reporting overheads and by transferring risk; a conceptual framework was proposed for the use of decentralized autonomous organization (DAO) [41]. However, it was not validated. Others argue that the construction supply chain cannot adopt blockchain and smart contracts unless it undergoes major transformations with regard to its business model and procurement arrangement [42]. The weaknesses and threats can be further mitigated by a better understanding of the role of project stakeholders and their responsibilities [42].

A peer-to-peer (P2P) system was proposed for collaboration [43], where construction stakeholders become authorized to participate as miners after receiving a digital certificate from institutions such as BuildingSMART. While the proposal takes advantage of blockchain's consensus algorithm for content curation in AEC, it needs more specific use cases and a demonstration of its applicability to a particular domain with real-world validation. A smart contract-based solution was proposed for quality acceptance in construction [44]. The research modeled the workflows based on a set of state transitions, taking advantage of the underlying blockchain protocols. The relationship between actors, their decision, and how they enable state transitions were further formulated [45]. The method needs to be validated in future. A series of semi-structured interviews [46] revealed that smart contracts reduce the cost of contract signing and bidding, and they enhance the projects' capital flow issues. Opportunistic behaviors can be minimized on projects where smart contract use provides historical data with regard to project stakeholders and their participation [46]. Future work needs to focus on the integration of smart contract with new contractual models [47].

## 2.1 Applications of Smart Contracts for Payment Processing

The automation of contract administration and relevant project management workflows have been identified among the first viable application areas for smart contracts and blockchain technology in AEC [37,48,49]. This focus area has received increasing attention in the past few years and due to the technology's elimination of third-party intermediaries and the manual work involved in payment processing [50,51].

One study identified the increase in payment transparency and the quality of cost and schedule data as the key values added by smart contracts [15]. Another work explored the 'why' of blockchain for payments [17], arguing that payment automation can be achieved using alternative technologies but not reliably; blockchain-enabled smart contracts addresses two major shortcomings of current payment solutions: 1) centralized control mechanisms, and 2) lack of guaranteed execution [17]. If successful, smart contract-based payment solutions have benefits that go beyond improved cash flow; establishing a seamless payment system is the first step toward e-commerce applications in AEC [52].

Few studies have focused on the security of payments: a smart contract-based solution was introduced [53] to safeguard subcontractors and contractors from the insolvency of the project owners. At the beginning of a cycle, the contractor requests a blockage of funds stored in smart contract for a period of 30 days, to be released afterwards upon owner's review of progress. This guarantees that funds will be available at the end of payment period; the workflows for submitting and reviewing payment applications, however, still follow traditional solutions and are susceptible to shortcomings caused by centralized control. A trust-less system was introduced for secure sharing of payment records among stakeholders and semi-automatic enforcement of interim payment terms [54].

Others have investigated the automation and decentralization of the workflows around payments: CryptoBIM [55] creates a shared, immutable, and distributed view of BIM on blockchain to facilitate the data transfer between BIM and smart contracts; such integration is key to successful adoption of smart contracts [23]. Another study [56] introduced a semi-autonomous method based on Hyperledger [57] that comprised two steps: the contractor manually submits the application for payment along with the supporting documents (e.g., quantity of material) to a consensus algorithm where each project participant acts as a node in the blockchain network; if all nodes approve the submission, it will be passed to a party for manual authorization. This work demonstrates the use of consensus algorithm for enhancing payment processing. However, it requires the stakeholders to run full nodes; the method can be subject to potential gatekeeping due to its reliance on intermediated workflows that are present in today's payment applications.

In a case study-based approach, a framework was developed for integrating smart contracts, sensing technology, and BIM for payment automation [58]. A notable contribution by this work was conditioning the smart contract execution on the real-world sensor feed. The smart contract, however, was developed by a third-party vendor and its design was not detailed. The solution was designed for elements that are installed on job sites, not applicable to all element types. A framework was introduced for automated billing by integrating 5D BIM and smart contracts [59], where payments between the client and the general contractor are executed using a billing model stored in a common data environment [60]; the framework will be implemented and validated in future.

The integration between cash and product flows is another area that can benefit from smart contract use [61]. Integration is key to automation [62], and bringing cash and product flows closer can streamline payments. A business model was conceptualized to promote the integration of flows within the context of Swedish construction industry and for use by logistics consultants [63]. It was demonstrated that crypto assets (crypto currencies and crypto tokens) can be used in the place of fiat to enhance the atomicity and granularity of the integration between the flow of cash and products [64].

## 2.2 Points of Departure

There is a need for a smart contract-based solution that can automate the transition from progress data, enabled by reality capture technologies, to construction payments. Such automation needs to be decentralized, free of intermediated workflows, free of manual applications for payments, and capable of using reality capture data available in semantically-rich as-built BIMs. This is currently not possible.

The existing literature clearly demonstrates the power of smart contracts for enhancing construction payments. The applications of smart contracts, reviewed herein, eliminate some of the overhead around payment processing; however, much of the workflows still rely on intermediated processes (e.g., manually submitting, reviewing, and approving applications for payments and their supporting documents). These create opportunities for centralized control, denials of service, and hurt the guarantee of execution. Additionally, current applications of smart contract do not support the use of on-site reality capture.

The mere computerization of payment application is not the goal for smart contracts, as automation can be achieved using other alternatives. It is necessary to tailor the design of the smart contract solution based on the properties of the industry problem [65] to take advantage of its key decentralization feature:

smart contracts' key value proposition is enabling *reliable* automation [66] and "verifiable confidence" [67] in payment data by eliminating intermediated workflows and by creating guarantees and predictability [17].

The broader research landscape has seen very few validations [68] and is mostly theoretical and not ready for use in real-world projects [69]. The real-world implementations is key to the technology's adoption among AEC practitioners [70], key stakeholders that currently have doubts about the automation capability of this technology [71]**.** Therefore**,** in addressing the limitations discussed above, this work focuses on real-world implementation of its proposed solution.

## 3    METHOD: AUTONOMOUS CONSTRUCTION PROGRESS PAYMENTS

Each construction progress payment can be viewed as a transition from product flow to cash flow (Fig. 1). Borrowing from the works of John Searle [72,73], the authors distinguish between the two types of facts involved in payment processing. These are respectively referred to as "brute facts" and "institutional facts". The former describes the status of construction work at job sites and the observations regarding the product flow; the latter describes the flow of cash, and its status depends on human agreements, the choice of observer, and their interpretation. In the context of construction payments, institutional facts comprise, among others, the interpretation of payment terms and contractual agreements. In this work, the terms "physical reality" and "reality" are used interchangeably to refer to the brute facts (i.e., product flow). The term "social reality" is used to refer to institutional facts (i.e., cash flow).

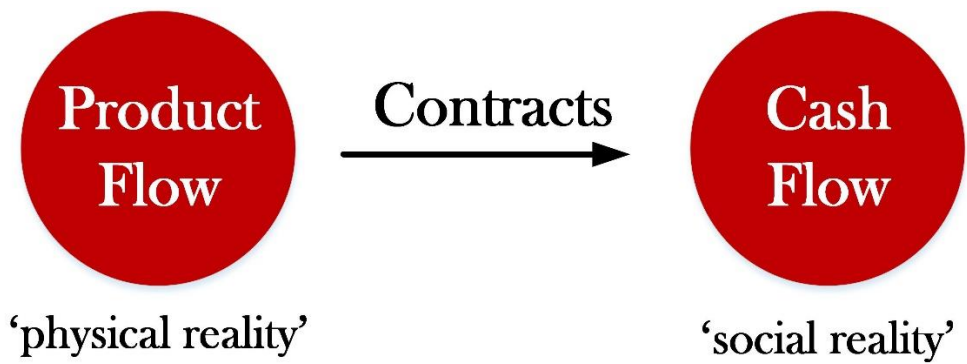

Fig. 1. Construction progress payment is a transition from product flow (physical reality) to cash flow (social reality)

Automating the transition from product to cash flow faces challenges that motivate the design of the proposed method:

For construction payments to be automated, different stages of payment workflows need to be computerized in the form of scripts and executed on computational resources. These stages include, among others, the analysis of on-site observations, the valuation of the completed work, and importantly the contract execution and the issuance of financial transactions. One or more of the project participants need to execute these computerized codes and broadcast the results (the valuation of the work and the resulting accounts receivable and payable) to others before transactions are executed by banks. Such automation is not reliable; this client-server architecture creates single points of failure, and the centralization and the skewed concentration of power exposes payment processing to potential denials of service, re-writing of the past data, and information asymmetry. Such automation, therefore, suffers from the same limitations of today's payment applications.

This work uses *Blockchain-enabled smart contract*s to eliminate this vulnerability; the smart contract use adds reliability to payment automation by guaranteeing the execution of contract terms and eliminating the centralized client-server model. A smart contract memorializes the project teams' contractual agreements in the form of a script that is executed in a decentralized way on blockchain: *all* blockchain miners (i.e., full nodes) *independently* execute the contract script, compute the results, and reach consensus regarding the financial transactions, accounts payable, and accounts receivable.

Smart contracts, however, are as reliable as the consensus algorithms governing their underlying blockchain. The design of these consensus mechanisms varies based on the choice of public, private, and consortium blockchains (i.e., private blockchains run by a pre-selected group of institutions). This work deploys its smart contract on public blockchains to ensure the reliability of payment automation: in private and consortium chains, project participants act as miners, execute the contract code, and reach consensus regarding the true state of physical and social realities; such uses of permissioned blockchains are susceptible to anti-trust risks [74,75], where participants can collude and jeopardize the decentralized nature of the consensus algorithm. The immutability, irreversibility, and guarantees of execution do not always hold.

This work deploys its smart contract on public blockchains, where project participants do not directly participate in the consensus algorithm and the execution of contract code; they cannot jeopardize its integrity. The Ethereum [76,77] and Bitcoin blockchain [78,79] are the two prominent permission-less chains with the highest hash rate and therefore highest security. The former was used for the development of the method due to its support of generic developments within the decentralized application ecosystem and its more powerful scripting language compared to the latter.

The application of smart contracts running on public blockchains, as discussed so far, decentralizes both the payment administration and the documentation of cash flow, providing reliable access to a project's *social reality*. However, the smart contract use alone cannot support payment automation due to the technology's *disconnect* from the real-world events *(physical reality)* and its inability in creating a link between the *off- and on-chain states*. This is a key challenge that limits the applicability of smart contracts in construction management applications.

To address this challenge, this work equips the smart contract with an *eye* in the real world (i.e., an *'oracle'*) using reality capture solutions. The term "reality capture" is used herein to refer to the use of sensing, machine intelligence, and BIM for respectively capturing, analyzing, and documenting the status of physical reality at construction job sites.

The importance of reality capture to payment automation is twofold: First, it makes it possible to condition the blockchain's state transitions on real-world events. Second, it introduces objectivity: the valuation of the work becomes independent of the observer, increasing the reliability of the input data provided to smart contract and ensuring consistency across projects and across each project's life cycle. This is an improvement to manually-reported progress reports which can jeopardize the reliability of payment automation due to their reliance on the role of individual observers and their relatively more subjective interpretation of brute facts.

Once successfully captured, the physical reality needs to be securely shared among project participants to create a common understanding of the brute facts, before it is used by smart contract to flow funds. This work uses the InterPlanetary File System (IPFS) [80] to create a distributed and content addressable file sharing solution. Content addressability, the key innovation of the IPFS, means that project information is referenced off- and on-chain by *what* it is (content) and not *where* it sits (location). Each data instance (e.g., an as-built BIM) has a unique identifier, calculated using cryptographic hash functions, that changes with each update to the data instance. This content addressability is crucial to the smart contract's operation in this work and allows for creating immutable and secure links to physical reality on the public blockchain (this is discussed in more detail in section 3.2). This approach contrasts with location-based data referencing, where project information is retrieved based on *where* it is stored on centralized data servers.

Fig. 2 illustrates how these four design elements*, reality capture technologies, IPFS, smart contract, and public ledgers* are incorporated to achieve payment automation, translating the progress at job sites (Fig. 2a, product flow) to construction progress payments (Fig. 2c, cash flow) without reliance on existing centralized, intermediated, and resource-intensive workflows such as invoice collection and application/certification for payments.

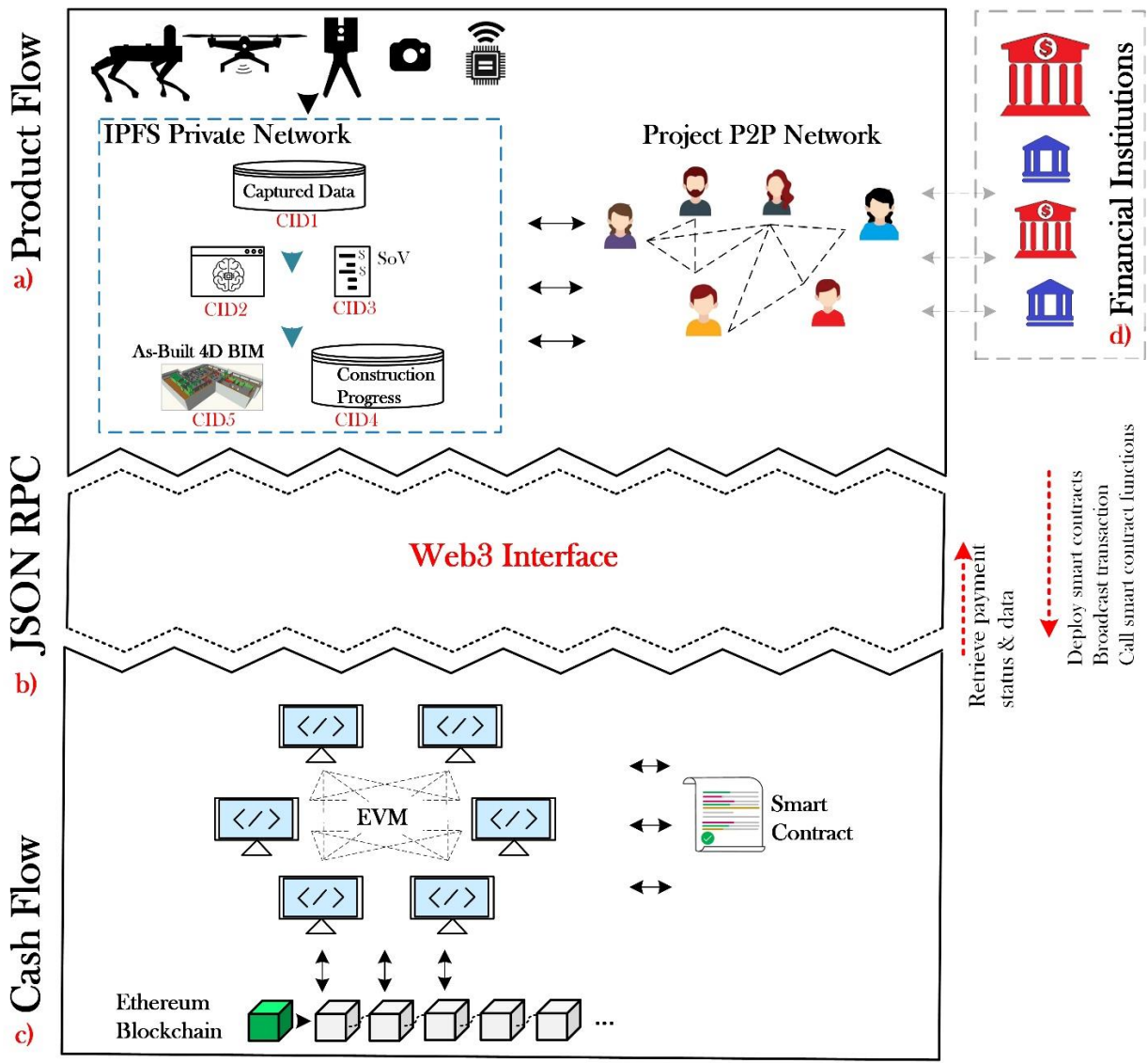

Fig. 2. Autonomous construction progress payment: a) product flow stored on private IPFS network; b) the JSON RPC interface facilitating the flow of information; and c) cash flow moderated by smart contract living on the Ethereum blockchain; d) financial institutions on the periphery

The following sections respectively provide more detail with respect to the design of the off- and on-chain components of the proposed method. The section 3.1 details how the product flow is captured and securely stored in a distributed manner. Section 3.2 discusses the design of the smart contract used for payment automation and how it uses progress data to achieve on-chain payment settlement.

### 3.1 Off-Chain Flow of Products

Accurate and timely understanding of changes in the physical reality (i.e., product flow) is the first step toward successful payment automation. In this work, construction progress is measured using reality capture solutions operated manually (e.g., laser scans) or mounted on robotic platforms (e.g., camera-equipped unmanned aerial vehicles). The captured data is analyzed using machine intelligence to arrive at the percentage completion data for various scopes of work. This is achieved either by identifying the components (e.g., columns, studs, insulation panels) or the state of progress for building elements (e.g., painted partition). Depending on the choice of machine intelligence algorithm, the progress results can vary in terms of its granularity, available per globally unique identifiers (GUID) or per a larger scope of work such as one or few building floors. The resulting progress data is automatically incorporated into as-built 3D/4D BIM. This creates an integrated data-driven approach through which job site physical reality can be directly used to value the work completed.

After the progress of the work is captured and analyzed, it is categorized and valued using the project's corresponding cost codes, making the product flow structured for use by smart contract. There are few challenges associated with this step: 1) the product flow cannot be stored on blockchain due to prohibitively expensive on-chain storage; 2) storing data on blockchain risks the integrity of project information due to the auditability of Ethereum public blockchain; 3) project participants need a means of referencing the same data throughout a project's life cycle without duplicating files across platforms and organizations; and 4) the product flow documentations used in payment processing need to support auditability and the reproducibility of the computations.

To address these barriers, this work proposes the use of *content-addressable* file sharing for off-chain storage of product flow and across the project P2P network. The key data including the schedule of value, as-built BIMs, progress data, and the scripts used for analyzing the status of construction work are stored on a private IPFS network (Fig. 3a).

IPFS breaks down files into blocks that are stored in key-value pairs in a Merkle directed acyclic graph (DAG). The Merkle DAG differs from Merkle tree by its lack of balancing requirement; its use of multi-hash checksum brings content addressability and tamper resistance: once a file is stored on IPFS, a unique content identifier (CID) is automatically generated for the file and can be used to retrieve it from the network. A file's CID is based on its SHA-256 hash and therefore changes with each update to the underlying file. For example, an as-built BIM used for the valuation of work automatically receives a new CID each time it is updated to incorporate new progress data.

The content addressability stems from the IPFS' ability to reference and retrieve files by 'what' they are and not by 'where' they are stored. All IPFS peers (project participants in the case of private IPFS) are connected to the same data structure explained above. Hence, 1) a CID generated on one node can be used by another node to retrieve the same file throughout the project's life cycle; and 2) A file will be recognized by the same CID regardless of who uploads the file, when it is uploaded, and how many times it is uploaded.

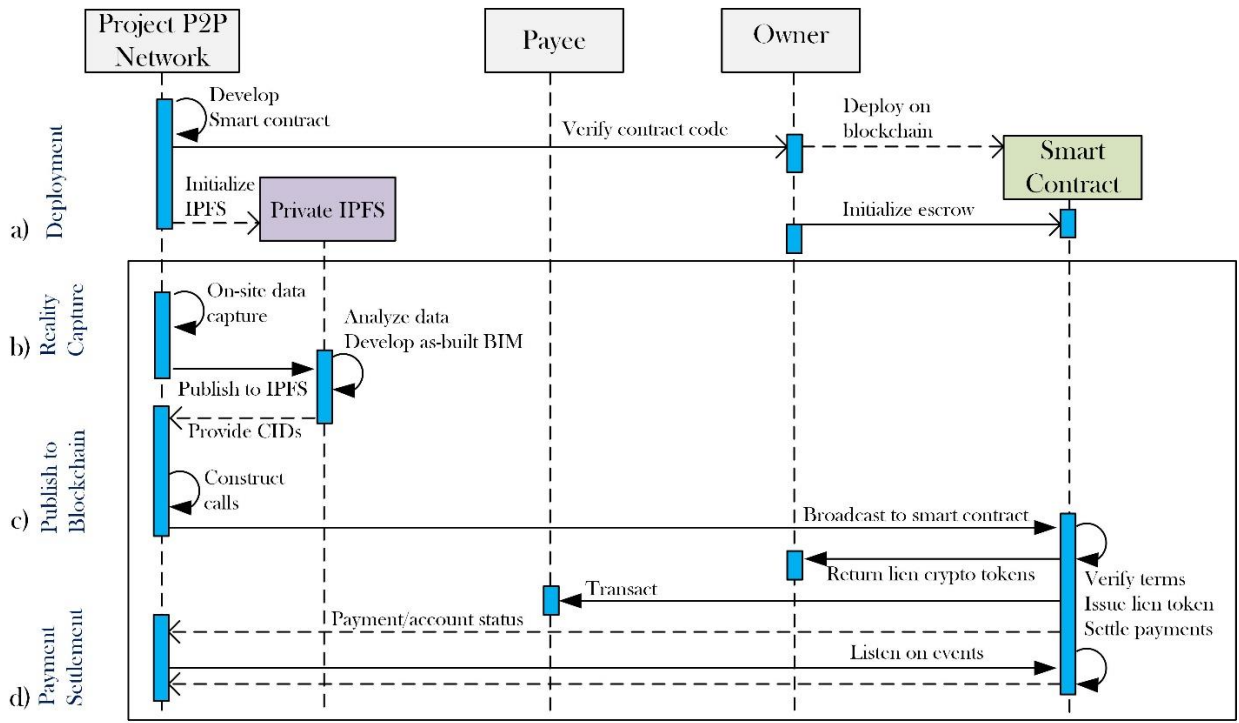

Fig. 3. The sequence diagram of the workflows involved in smart contract-based autonomous payments: a) smart contract deployment on blockchain; b) on-site reality capture; c) publishing progress data to blockchain; and d) on-chain payment settlement

The use of IPFS is critical to this work's off-chain documentation of product flow and its later use by smart contract: first, project participants use permanent CIDs (e.g., "QmXvTCjaxAW4YHnnQMePU1kahvJ1s8jEEuYRwPNppDUFXn") rather than locations on a project server (e.g., "https://cife.stanford.edu/TR233") to retrieve key data used in payment processing. This creates a common view of the product flow, eliminates duplications, and ensures consistency. Second, it bridges the off-chain and on-chain states; the CIDs corresponding to key project information used in payment processing are communicated to the smart contract and included in state transitions, recorded on the public ledger (Fig. 3b-c). These references cannot be interpreted by a public viewer of the Ethereum blockchain who is not part of the private IPFS network, but they enable project participants to retrieve the raw data used in payment processing. Third, content addressability protects payment automation from potential attacks due to its tamper resistance, preventing stakeholders to re-write the past, and its relatively lower susceptibility to network down time, preventing the denial-of-service attacks.

## 3.2 On-Chain Flow of Cash

The on-chain management of cash flow needs to accomplish two objectives: 1) settle payments between project participants in accordance to off-chain flow of products, and 2) transfer lien rights alongside payments. The mechanics lien is a contractor's claim on the property, and it can be filed for work that is not compensated.

Achieving these two objectives in an autonomous and decentralized payment system is challenging because 1) the financial institutions (e.g., banks) have a peripheral role in the proposed method (Fig. 2d) and do not directly manage the flow of cash; money is an institutional fact and needs the backing of centralized trusted third parties, the bodies whose role is disintermediated; 2) in preparing today's payment applications, general contractors collect conditional lien waivers (a written agreement by a subcontractor to waive the right to the property upon the receipt of payments for submitted invoices); these workflows are not available in a decentralized approach that eliminates applications for payment.

These challenges are addressed by introducing a smart contract that autonomously settles payments and transfers lien rights respectively using the Ethereum's native cryptocurrency (Ether, "ETH") and a non-fungible token. Tokens (crypto tokens) represent the right to a utility or asset (e.g., monetary value and ownership). This is also known as the smart property feature [81], with tokens giving their owners the control of an asset. The two most common standards used in defining crypto tokens are the ERC-20 [82] and ERC-721 [83], respectively used to define fungible and non-fungible tokens. Each lien right is unique and corresponds to a particular scope of work, as documented in the off-chain product flow records. Therefore, this work utilizes the ERC-721 standard to create the non-fungible "LIEN" token to represent the lien rights to a property; the ownership of the token, and hence the right to underlying physical asset, is managed by the smart contract and recorded on the Ethereum blockchain.

The smart contract introduced herein (Fig. 4a) is deployed on the Ethereum virtual machine (EVM) by the project owner. The smart contract account address is both public and payable, and it acts as an escrow account (Fig. 3a) which conditions its outputs, the flow of cash and the transfer of lien rights, based on the input it receives regarding project's physical reality (Fig. 4b-c). Contract accounts cannot initiate transactions on their own, and the externally owned accounts (EOA), controlled by project participants, need to trigger smart contract functions. The method's off-chain component (detailed in section 3.1) identifies the incremental construction progress, as analyzed by the reality capture solution (Fig. 3b), and it values the work using the project's schedule of values, the building elements' semantics and geometry (retrieved from BIM), and the percentage completion data. This process is independent of the level of granularity; it can be high for projects where as-built modeling is performed at a building element level and low for projects where the progress evaluation is performed as a percentage competition for the work on multiple building elements (e.g., electrical work 80% complete).

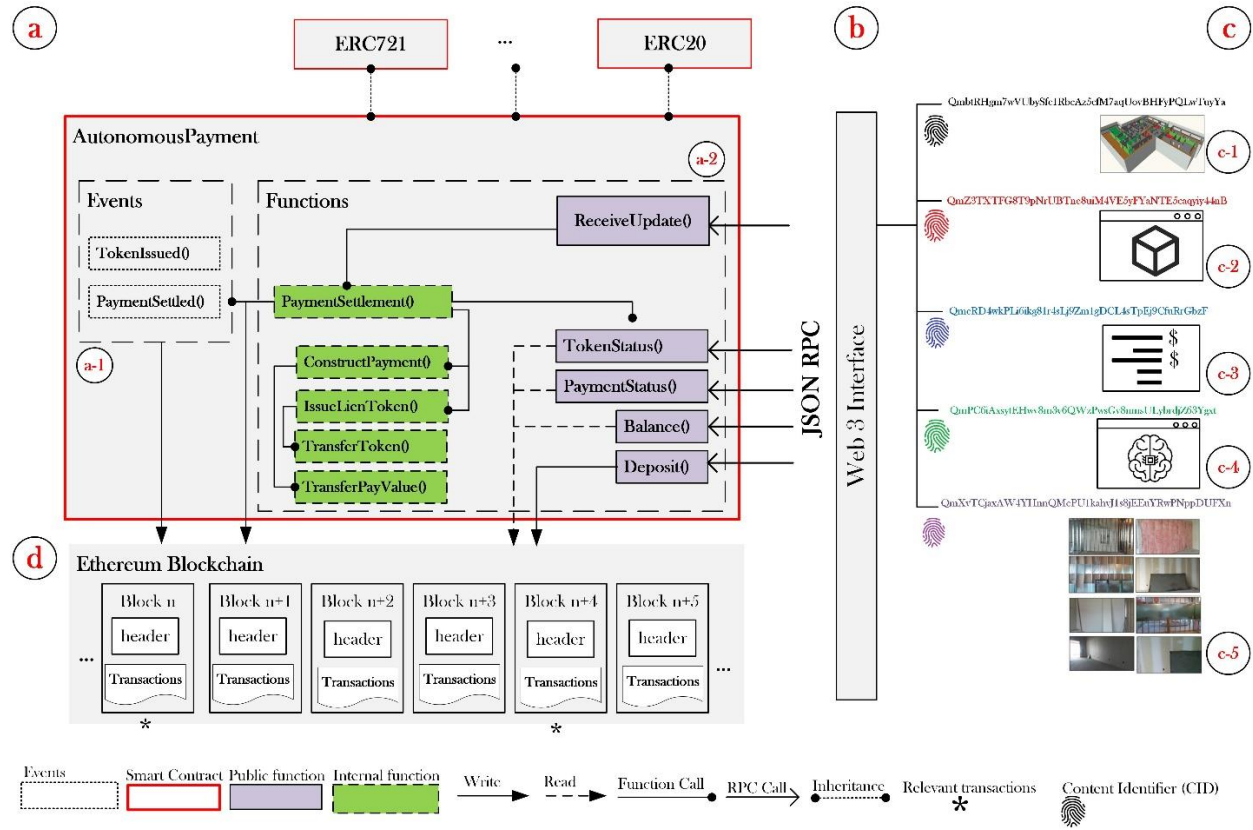

Fig. 4. The design of smart contract for autonomous payments and its interactions with project P2P network: a) the contract structure, b) JSON RPC, c) the CIDs generated on the private IPFS network; d) transactions written to the Ethereum blockchain

The off-chain client constructs a transaction that is broadcasted to smart contract (Fig. 3c) via a JSON-encoded remote procedure call (RPC) (Fig. 2b), communicating the valuation of the work, the public key of the subcontractors performing the valued work, and the CIDs of input data used for the valuation of incremental progress; the latter includes CIDs for the list of elements and their GUIDs, as-built BIM, schedule of values, raw progress data, and the automated solution used for progress evaluation (Fig. 4c).

This data is broadcasted to the smart contract's public function, *ReceiveUpdate* (Fig. 4a-2) which ensures the information is correctly formatted before it triggers internal functions to initiate on-chain payment settlement.

The function checks the address of the node publishing the updates against the addresses of the nodes in the project's P2P network; it updates the smart contract parameters by writing the changes to the underlying Ethereum blockchain, keeping track of processed payments and their corresponding records of elements, payees' public key addresses, issued transactions, and lien tokens. This is crucial to avoiding double payments. Alternatively, event logs can be utilized to achieve similar functionality. In the latter case, an off-chain client, executed by project P2P network, needs to accumulate the list of payments. The *ReceiveUpdate* function triggers internal functions that handle payment settlement.

The internal functions cannot be triggered from outside the smart contract, and they include *PaymentSettlement*, *ConstructPayment*, *TransferPayValue*, *IssueLienToken*, and *TransferToken* (Fig. 4a). Their logic and parameters are updated and digitally verified by project stakeholders each time there is an amendment to the project's contractual agreements.

Once the public function *ReceiveUpdate* triggers the internal function *PaymentSettlement*, the latter verifies the smart contract account's balance and initiates the internal functions *ConstructPayment* and *IssueLienToken* to construct the transaction that compensates the payee and to mint a new ERC-721 LIEN token representing the lien rights associated with the valued work (Fig. 3d). The *TransferPayValue* transfers the payment in Ether and to the contractors responsible for the work, as specified by *ConstructPayment*.

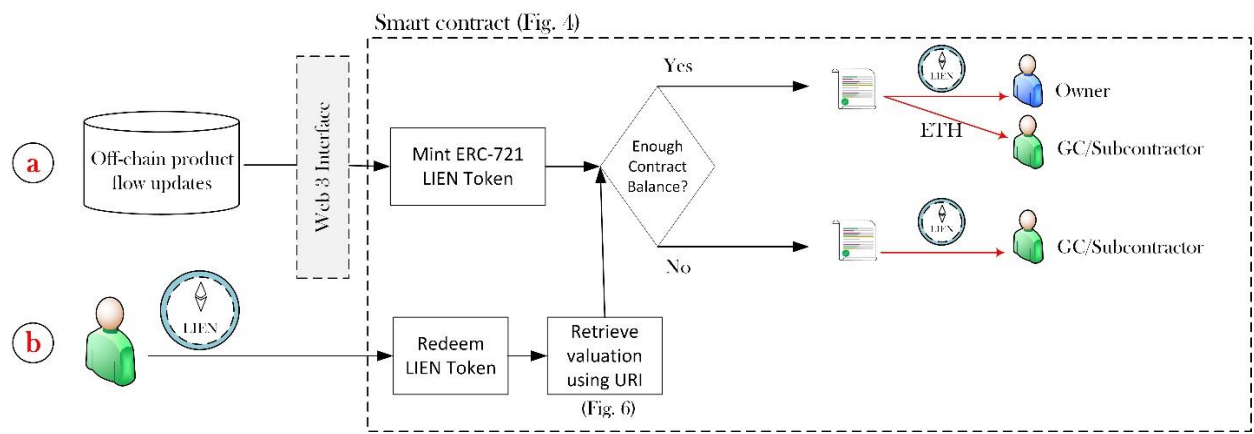

Fig. 5. The application of ERC-721 for the transfer of lien rights: a) the issuance of LIEN token and payment settlement upon the receipt of new progress updates; b) the redemption of LIEN token by a subcontractor, receiving payment by returning the collateral

The payment metadata, including references to the building elements and the payment value, are incorporated into the lien token. This is achieved by populating the ERC-721 token's uniform resource identifier (URI) with the CID corresponding to the building elements for which the right is transferred. Upon payment settlement, the corresponding lien token is transferred to the owner using *TransferToken* (Fig. 5). If the balance is not enough, the token will be automatically transferred to the contractors responsible for the work (Fig. 5a), giving the contractor right to the work described in the product flow records and available via CIDs. The ERC-721 token can be sent to the smart contract address to be automatically redeemed for its value once the contract address has enough balance (Fig. 5b): the valuation of an ERC-721 token, corresponding to the product flow updates that initiated the payments, can be retrieved using its URI and by reading from the smart contract's storage.

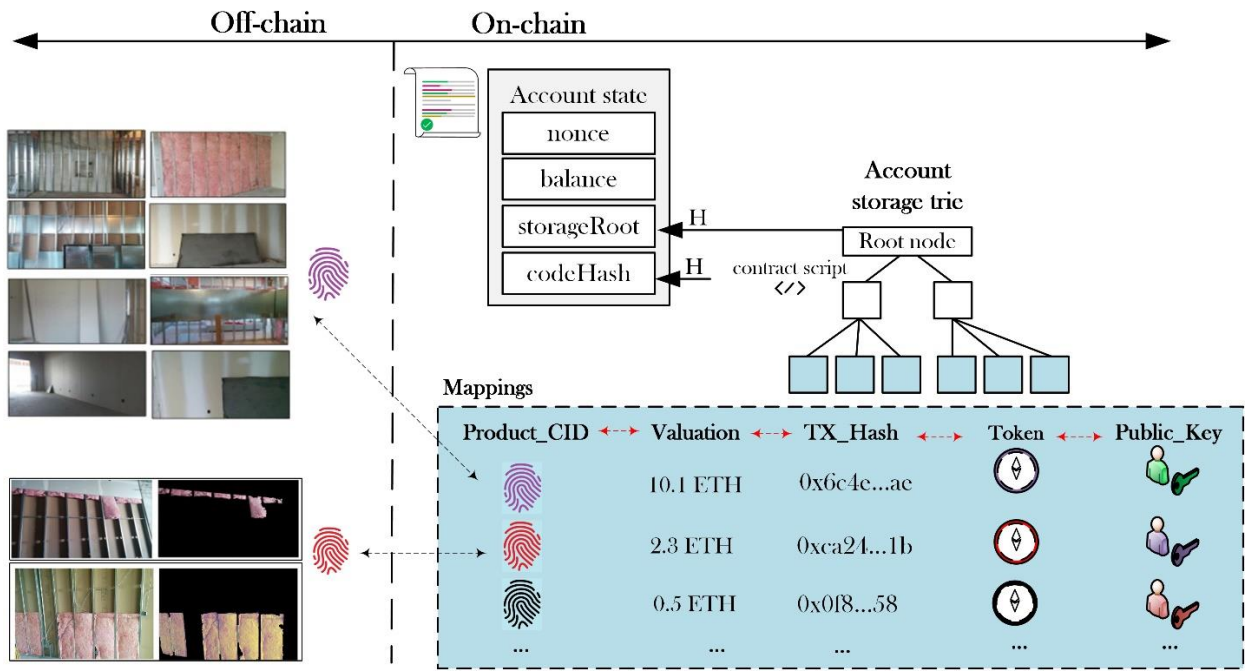

Fig. 6. The smart contract's on-chain documentation of cash and product flows creates a series of mappings between product flow (CIDs and the valuation) and cash flow (corresponding transactions on blockchain, corresponding LIEN token, and the public key of payee)

Fig. 6 illustrates the blockchain's data architectures that support the retrieval of such information; the smart contract directly writes to the Ethereum blockchain (Fig. 5d), and the contract's parameters can be read from the Account Storage Trie (accessible from State Trie under each block header). This creates a series of mappings between instances of product CIDs, valuations, LIEN tokens, public key addresses of payees, and the corresponding transactions, among others. These mappings allow for the retrieval of cash and product flow information when either of the above parameters are known. For example, analyzing a transaction that compensates a contractor, the owner can retrieve the corresponding product flow data. An ERC-721 token's URI can be used to retrieve the corresponding progress data and the valuation of that scope of work. These mappings also create a link between off- and on-chain records because CIDs provide permanent references to data stored on the project's IPFS network.

## 4 EXPERIMENTAL SETUP AND RESULTS

The proposed smart contract-based solution was implemented in Solidity v0.6.2. The web3.py and Infura were respectively used to set up the RPC module and to enable queries about the status of Ethereum blockchain.

The method was used for processing progress payments on two commercial construction sites. The transactions were processed on Ethereum test network, and a private IPFS network was created among the project participants including the owner, the general contractor, and the subcontractors. Table 1 lists the description of each project, the number of trades for which the payments were processed, and the type of building elements covered in the reality captures. Project A is in the state of Ontario (Canada), and it was visited for data collection over a period of 4 weeks. Project B is in the state of California (United States), and data capture was performed at the site for a period of 5 months.

**Table 1. The two commercial construction sites where the smart contract-based payment system was deployed**

| Project | Location | number of floors inspected | building floor area (sq.ft) | number of subcontractors | Building elements |
|---|---|---|---|---|---|
| A | Ontario (Canada) | 1 | 40,000 | 4 | Partitions |
| B | California (USA) | 4 | 200,000 | 3 | HVAC, Plumbing, Partitions |

The data capture at Project A was performed using a camera-equipped unmanned aerial vehicle (UAV) (Fig. 7a), resulting in digital images (Fig. 7b); a laser scanner mounted on an unmanned ground vehicle (UGV) (Fig. 7c) was used to capture the state of construction progress at Project B (Fig. 7d).

Table 2 and Table 3 list the tests that were designed to evaluate the feasibility of using the proposed method. The tests performed at Project A (test 1) and Project B (test 2) use different reality capture solutions, sensors, data analytics tools, and data formats; additionally, the first test has more granular product flow (Table 3) with progress data available per 104 inspected building elements with connection to BIM. The second test's product flow, on the other hand, is low in granularity and provides aggregate percentage completion. These disparities are intended to help better evaluate the generalizability of the proposed method and its applicability in commons progress monitoring practices.

**Table 2. The tests designed to evaluate the feasibility and performance of the smart contract-based system**

| Test | Project* | Data capture | Sensor | Reality capture methods | | Data formats | |
|---|---|---|---|---|---|---|---|
| | | | | Progress assessment | As-built 4D BIM generation | BIM | schedule |
| 1 | A | UAV (Fig. 7a) | Camera (Fig. 7b) | [84,85] | [86,87] | IFC | IFC |
| 2 | B | UGV (Fig. 7c) | Laser scans (Fig. 7d) | Outside vendor | - | .rvt, .nwd | XML |

* See Table 1 for project details

**Table 3. The payment frequency and the level of granularity for the performed tests**

| Test | Payment frequency | Product flow granularity | Duration of site visits (weeks) |
|---|---|---|---|
| 1 | Bi-weekly | High* | 4 |
| 3 | Monthly | Low** | 20 |

* Progress data collected and analyzed per GUID

** Aggregate progress data reported for all inspected building elements

In both tests, the general contractor deployed the robot on site; this involves an individual mounting the reality capture technology on the robotic platform and returning the robot to a safe location after data collection. The raw progress data was manually extracted from the robot's memory card and sent to the authors (test 1) and an outside vendor (test 2).

In test 1, UAV-captured digital photos were passed to an automated computer vision-based solution [84,85] to generate the states of progress for indoor partitions. The progress data was incorporated into industry foundation classes (IFC)-based 4D BIMs using an automated method [86,87]. The data analytics and in-BIM documentation of progress did not require human involvement. In test 2, the UGV-captured point clouds were processed by an outside vendor. The algorithms used by the vendor and their degree of automation are not known to the authors, but the process is believed to require significant human involvement. As-built BIMs were not created in test 2.

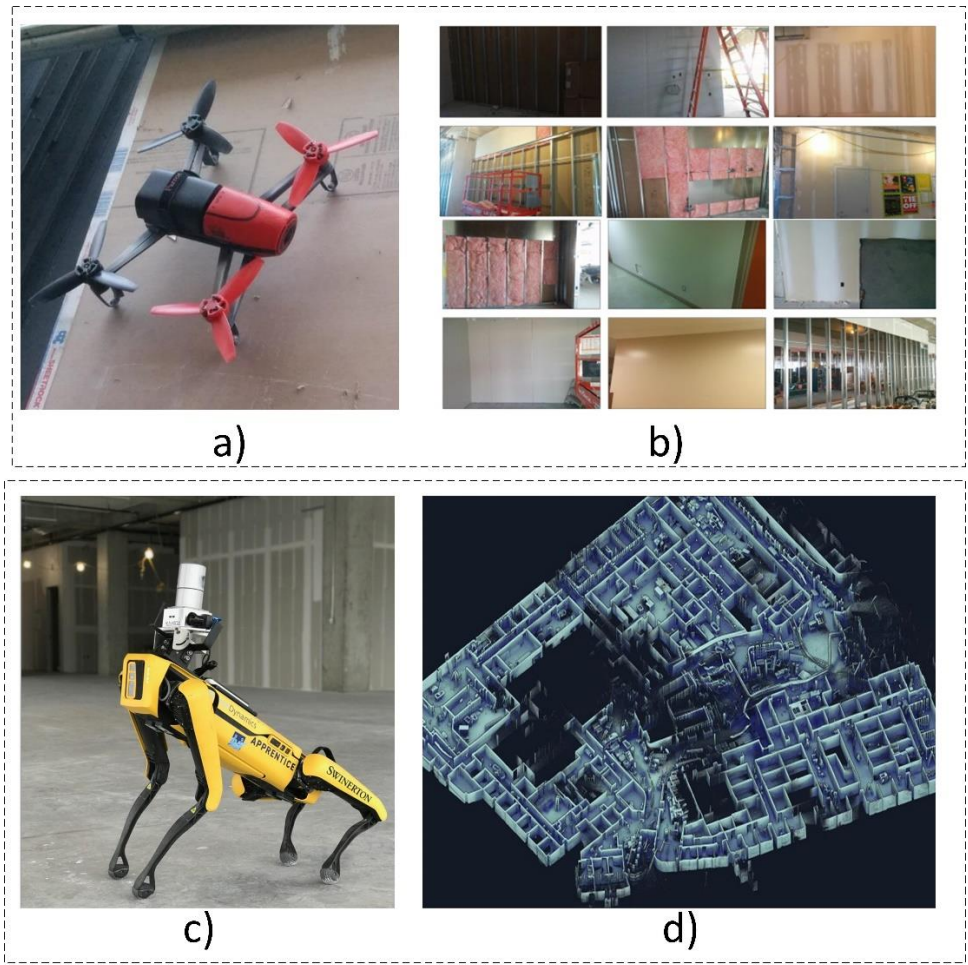

Fig. 7. Reality capture tools used in the tests: a) quadrotor UAV used for image and video capture; b) UAV-captured images; c) laser scanner mounted on a UGV; d) UGV-captured point cloud of an inspected building floor (Fig. 7c-d courtesy of Swinerton)

In both tests, the progress data, the as-built BIM (when available), and the supporting files used in the valuation of the work were uploaded to the private IPFS network, where the CIDs are automatically constructed. One of the P2P nodes communicates the CIDs to the smart contract via the RPC module; the smart contract operation (Fig. 3d) did not involve human involvement.

Table 4 lists the results of the performed tests, including the processing time per each payment cycle, the accuracy of both remote sensing and the proposed smart contract-based solution, and the number of payment cycles. The progress assessments and the method's outputs including the transactions were manually verified to assess the accuracy. The accuracy of the remote sensing solution is defined as the percentage of elements for which the state of progress was accurately identified (i.e., reported as-built condition matches the on-site condition, manually verified by the authors). The reality capture accuracy is not available for test 2 because the outside vendor does not manually verify the on-site condition, nor they publicly share the accuracy of their platform. As for the payment processing, a payment is considered accurate if the monetary value transferred by smart contract, the recipient, and the incorporated metadata (e.g., CIDs) match the ground truth. All transactions were manually verified, and no issues were observed with the accuracy of the smart contract's payment settlement. In addition, the payments were successfully settled for the entire scope of work covered in the progress reports.

**Table 4. The results of the performed tests**

| Test | Number of payment cycles | Number of blockchain transactions | Average processing time per payment cycle | | Average accuracy | |
|---|---|---|---|---|---|---|
| | | | Reality capture* | Payment processing** | Reality capture | Payment processing |
| 1 | 2 | 10 | 9 minutes | 5 minutes | 95% | 100% |
| 2 | 5 | 20 | 5 hours*** | 4 minutes | - | 100% |

\* Automated detection of progress and model updating (excludes data collection time)

\*\* The construction of calls to smart contract, smart contract execution, and on-chain payment settlement (12 block confirmations)

\*\*\* Reported by outside vendor

The payment processing times reported per cycle (Table 4) are comparable across all tests because the smart contract-based solution uses the output of the remote sensing solution and is blind to the inner workings of data analytics tools employed for progress assessment. The processing time for the reality capture, on the other hand, varies significantly based on the choice of sensor and progress tracking solution. Computer vision-based analysis of point clouds (test 2) is more time consuming than digital images (test 1). The processing times reported in Table 4 do not include the data collection time and data preparation.

## 5 DISCUSSION

Bridging the divide between physical and social realities, as demonstrated in this work, is a necessary step toward payment automation. The proposed method successfully arrived from the observations of job site conditions (physical reality) to progress payments (social reality) without relying on applications for payments and other traditional workflows. This work decentralizes different stages of payment processing to a large extent. However, not all steps can be decentralized, nor it is necessarily favorable to achieve full decentralization.

The smart contract allows for a decentralized execution of contractual agreements in an automated system, eliminating the need for server-customer model; some steps involved in preparing the inputs and analyzing the outputs still require human involvement. This is particularly the case for the on-site deployment of robotic reality capture tools and the transfer of captured reality from data collection tools to analytics platforms. These stages are performed by few individuals and are centralized. Automating these steps is possible, but it does not necessarily enhance the reliability of payment automation. Human involvement can ensure the quality of on-site data collection and timely adjustments in the case of faulty data capture or malfunctions.

While the proposed smart-contract-based solution was demonstrated to be highly accurate (Table 4), the accuracy of its input data (outputted by reality capture technologies) is of paramount importance to seamless and reliable payments. For example, in test 1 (Table 4), the state of progress was inaccurately detected for 5% of the building elements (e.g., a painted partition classified as a plastered partition). The implication was that ~5% of payments, those associated with misclassified elements, were either delayed or processed sooner than expected.

This accuracy can still be higher than manually reported progress data, subjective to the interpretation of the inspector. Additionally, the discrepancy does not affect the total valuation of the work, but the timing of the 5% of payments for which the progress was misclassified. Depending on the progress data's level of granularity, the payments are documented per building element or cost code. In neither case, the percentage competition can be higher than 100%. Therefore, the smart contract's future payments will be respectively lower and higher than ground truth if the previous payments overcompensated or undercompensated a contractor.

The choice of reality capture also affects the scope of work that can be compensated in a single transaction. In the first test, the method processed payments at element level (e.g., framing completed for a particular wall). However, this was not possible in test 2 due to the progress reporting format of the outside vendor. The lower bound for the duration of the payment cycles seems to be limited by the reality capture practice and not the smart contract; the latter can process payments as frequently as data becomes available with respect to incremental construction progress. Taking into consideration the processing times (Table 4), the duration of robotic data capture at sites, and the data preparation times, the entire process can span few days; from a technical standpoint, payments can be processed within few days after on-site observations. The implications are significant for a cash-poor industry where it takes engineering and construction firms around 3 months to receive payment [88], with industry performance degrading over time [89].

Increased transparency is a key benefit that directly translates to the three desirable characteristics of smart contracts, as outlined in Nick Szabo's pioneering works on the concept [90-92]: 1) observability,

2) privity, and 3) verifiability. In a traditional payment system, the subcontractors cannot easily identify the source of payment delay, whereas in the smart contract-based method the status of payments and the completed workflows can be directly inquired from the contract account for each scope of work and without reliance on gatekeepers. This increases observability ex post. The availability of past interactions increases trust development [93] and hence increases observability ex ante.

The privity is ensured due to the elimination of third parties. The sensitive supply chain cash flow data, for example, is encapsulated from external financial institutions as they are pushed to the periphery (Fig. 2d). The verifiability feature is enabled via integration. The off-chain data storage and on-chain referencing enables project participants to verify past transactions and in relationship with the data captured on the job site, the script used for analyzing the data, and the schedule of values used for valuation. All smart contract's computations and the resulting cash flow can be re-computed and audited with permanent links to the input data. This creates trust by design and in an and inherently trust-less environment of competing business objectives.

# 6  CONCLUSION

The construction industry has been continuously moving toward digitalization, a trend that is significantly accelerated due to the strains imposed on its supply chain by the coronavirus pandemic [94]. Addressing the inefficiencies of the payment system is key to a successful transition toward the next normal.

Payment autonomy calls for a move away from today's manual and heavily intermediated workflows around payments including the preparation, review, approval, and execution of payment applications. Smart contracts, enabled by blockchain technology, are argued to address these limitations due to two key features: decentralization and guaranteed execution.

Despite its potential advantages, the technology's adoption in the payment systems faces a major barrier: a lack of a link between on-chain payment settlement and off-chain physical reality. The authors argue for an integration between digitized progress data, enabled by reality capture technologies and BIM-based progress monitoring, and smart contract protocols, not only to increase the feasibility of smart contracts application but to also increase the utilization of the semantically rich progress data captured at today's job sites.

This work proposed a smart contract-based solution for autonomously transitioning from on-site observations to progress payments. In this novel solution, stakeholders do not submit applications for payments but instead are compensated each time the on-site observations, as analyzed by machine intelligence, indicate an incremental progress. The application of the proposed method in two commercial construction projects demonstrated the feasibility of using smart contracts for accurate payment processing.

The accuracy of payment system was observed to be dependent on both the smart contract and the reality capture technique.

Smart contract-based payments are still in early stages of development and far from mass adoption. Future work must focus on the development of standardized and formally verified smart contracts, the fusion of various data sources for increased accuracy of product flow, and the development of hybrid models that allow for simultaneous on- and off-chain payment settlement.

# 7 ACKNOWLEDGEMENT

This work is financially supported by the Center for Integrated Facility Engineering (CIFE) at Stanford University (grants 2020-09 and 2018-06). The authors are grateful to Swinerton, PMX Construction, Perkins+Will, Inc., for their support during the data collection phase and granting access to project data. The first author extends his gratitude to Eric Law and Tristen Magallanes (Swinerton); Dr. Kincho Law, Dr. Michael Lepech, Dr. Forest Flager, Alissa Cooperman, Tulika Majumdar, Parisa Nikkhoo, and Yujin Lee (Stanford University); and Dr. Brenda McCabe and Pouya Zangeneh (University of Toronto) for their immense role in his PhD journey and for their invaluable intellectual companionship.